\documentclass{article}
\usepackage{spconf,amsmath,graphicx}
\usepackage{xcolor}
\usepackage{slashbox}
\usepackage{multirow}
\usepackage{pifont}
\usepackage{hyperref}

\title{Minimum Bayes Risk Training for End-to-End Speaker-attributed ASR}
\name{Naoyuki Kanda, Zhong Meng, Liang Lu, Yashesh Gaur, Xiaofei Wang, Zhuo Chen, Takuya Yoshioka}
\address{Microsoft Corp., USA}
\begin{document}
\ninept
\maketitle
\begin{abstract}
Recently, an end-to-end speaker-attributed automatic speech recognition (E2E SA-ASR) model was
proposed as a joint model of speaker counting, speech recognition and speaker identification
for monaural overlapped speech.
In the previous study, the model parameters were trained based on 
the speaker-attributed maximum mutual information (SA-MMI) criterion, 
with which 
the joint posterior probability for multi-talker transcription and speaker identification
are maximized over training data.
Although SA-MMI training showed promising results for overlapped speech consisting of various numbers of speakers,
the training criterion was not directly linked to the final 
evaluation metric, i.e., speaker-attributed word error rate (SA-WER).
In this paper, we propose
a speaker-attributed minimum Bayes risk (SA-MBR) training method
where the parameters are trained to directly minimize
the expected SA-WER over the training data.
Experiments using the LibriSpeech corpus show that
the proposed
SA-MBR training reduces the SA-WER by 9.0 \% relative compared with the SA-MMI-trained model.\footnote{The evaluation data used in our experiments can be found in \url{https://github.com/NaoyukiKanda/LibriSpeechMix}.}
\end{abstract}

\section{Introduction}

Speaker-attributed automatic speech recognition (SA-ASR)
from overlapped speech
has been an active research area
for meeting transcription \cite{fiscus2007rich,janin2003icsi,carletta2005ami}.
It requires to count the number of speakers,
transcribe utterances that are sometimes overlapped, and also diarize or identify the speaker of 
each utterance.
While significant progress has been made especially for multi-microphone settings (e.g., \cite{yoshioka2019advances}), 
SA-ASR remains very challenging when we can only
access monaural audio. 

A substantial amount of research has been 
conducted 
to achieve the goal of SA-ASR. 
One approach is applying speech separation (e.g., \cite{hershey2016deep,yu2017permutation})
before ASR and speaker diarization/identification. 
However, a speech separation module is often designed with
a signal-level criterion, 
which is not necessarily optimal for 
succeeding 
modules. 
To overcome this suboptimality, 
researchers have investigated approaches for jointly 
modeling multiple modules. 
For example, there are a number of studies  
concerning joint modeling of 
speech separation and ASR
(e.g., \cite{yu2017recognizing,seki2018purely,kanda2019acoustic,kanda2019auxiliary}).
Several methods were also proposed for integrating speaker identification and 
speech separation~\cite{wang2019speech,von2019all}. 
However, these studies focused on 
the combination of only a subset of the modules needed for SA-ASR.

Only a limited studies have tackled the
joint modeling of {\it all} the necessary modules for SA-ASR.
\cite{el2019joint} proposed to generate transcriptions for
different speakers interleaved by speaker role tags 
to recognize doctor-patient conversations.
In \cite{mao2020speech}, the authors applied a 
similar technique 
where multiple utterances 
are interleaved with 
speaker identity tags instead of speaker role tags.
However,
these methods 
are difficult to extend to  
an arbitrary number of speakers
because
the speaker roles or speaker identity tags
are determined and fixed in the training.
\cite{kanda2019simultaneous} proposed a
joint decoding framework for
overlapped speech recognition and speaker diarization,
where 
speaker embedding estimation and target-speaker ASR were performed 
alternately.
While their formulation is applicable to any number of speakers, 
the method was actually implemented and evaluated
in a way that could be used only for the two-speaker case, 
as target-speaker ASR was performed by using an auxiliary output branch
representing a single interference speaker \cite{kanda2019auxiliary}. 

Recently, an end-to-end (E2E) SA-ASR model was proposed
as a joint model of speaker counting, speech recognition, 
and speaker identification for monaural (possibly) overlapped speech \cite{kanda2020joint,kanda2020investigation}.
The E2E SA-ASR model has the advantage that it can
recognize overlapped speech of any number of speakers 
while 
identifying the speaker of each utterance 
from an arbitrary number of registered speakers.
It was trained 
based on the speaker-attributed maximum mutual information (SA-MMI)
criterion, by which
the joint probability for multi-talker speech recognition and speaker identification is maximized over training data.
Based on the SA-MMI training,
the E2E SA-ASR model achieved a significantly lower 
speaker-attributed word error rate (SA-WER) than a system
 that separately performs overlapped speech recognition and speaker
identification.
However, 
the training criterion was still 
not directly linked to the final evaluation metric, i.e., SA-WER.
As a result, considerable degradation of SA-WER was still observed for overlapped speech compared with non-overlapped speech, especially when the number of overlapped speakers were large.

In this paper,
to further improve the E2E SA-ASR model,
we propose a new training method, 
called speaker-attributed minimum Bayes risk (SA-MBR) training.
In SA-MBR training,
the entire network parameters are trained to 
minimize the expected SA-WER over the training data.
Note that there has been a lot of studies
on MBR training for the conventional (i.e. single-speaker, speaker-agnostic) 
ASR,
such as the one for hybrid ASR \cite{vesely2013sequence,su2013error,kanda2018lattice},
connectionist temporal classification \cite{sak2015acoustic,kanda2017minimum},
recurrent neural network transducers \cite{weng2019minimum,guo2020efficient},
and
the attention encoder decoder-based ASR \cite{prabhavalkar2018minimum,weng2018improving}.
However, to the best of our knowledge,
this is the first work that directly minimizes
the SA-WER in the joint framework.
We show that the proposed SA-MBR training achieves 
 significantly  better  SA-WER
over the model based on SA-MMI training.

\section{Review: E2E SA-ASR}
\label{sec:e2e-sa-asr}
\subsection{Overview}

In this section, we review the E2E SA-ASR model
proposed in \cite{kanda2020joint}.
The goal 
 is 
 to estimate a
multi-speaker transcription $Y=\{y_1,...,y_N\}$
and
the speaker identity of each token $S=\{s_1,...,s_N\}$ 
given
acoustic input 
$X=\{x_1,...,x_T\}$
 and a speaker inventory $D=\{d_1,...,d_K\}$.
 Here,
$N$ 
is the number of the output tokens,
$T$ is the number of the input frames,
and $K$ is the number of the speaker profiles 
(e.g., d-vectors \cite{variani2014deep}) 
in the inventory $D$.
Following the idea of 
serialized output training (SOT) \cite{kanda2020sot}, 
the multi-speaker transcription $Y$ is represented by concatenating each
speaker's transcription
interleaved by a special symbol 
$\langle sc\rangle$
representing the speaker change.

In the E2E SA-ASR modeling,
it is assumed that
 the profiles of all
the speakers involved in the input speech 
are included in $D$. 
Note that, as long as this assumption holds,
the speaker inventory may include 
 irrelevant speakers' profiles.

\subsection{Model architecture}

The E2E SA-ASR model can be decomposed to 
the {\it ASR block} and {\it speaker identification block},
which are interdependent. 
Here, we will briefly review the model architecture.
Interested readers can refer to \cite{kanda2020joint} for more details.

The ASR block is similar to 
the conventional attention encoder-decoder-based ASR 
and represented as follows.
 \begin{align}
 H^{enc} &=\{h^{enc}_1,...,h^{enc}_T\}={\rm AsrEncoder}(X).  \label{eq:enc}  \\
 u_n &={\rm DecoderRNN}(y_{n-1}, c_{n-1}, u_{n-1}).\label{eq:att2}   \\
 c_n, \alpha_n &= {\rm Attention}(u_n, \alpha_{n-1}, H^{enc}), \label{eq:att} \\
o_n &= {\rm DecoderOut}(c_n,u_n,\bar{d}_n)  \label{eq:asrout}
 \end{align}
Given the acoustic input $X$, 
an AsrEncoder module firstly converts $X$ 
into a sequence, $H^{enc}$, of embeddings for ASR (Eq. \eqref{eq:enc}).
At each decoder step $n$, 
DecoderRNN module updates the decoder state $u_n$ given 
the previous token $y_{n-1}$, previous context vector $c_{n-1}$, and 
previous decoder state $u_{n-1}$ (Eq. \eqref{eq:att2}).
Then, Attention module 
generates 
attention weight $\alpha_n =\{\alpha_{n,1},...,\alpha_{n,T}\}$
and 
the context vector $c_n$ as a weighted sum of
$H^{enc}$
(Eq. \eqref{eq:att}). 
Finally, DecoderOut module
calculates
the output distribution $o_n$ 
 given
$c_n$,
$u_n$,
and 
the weighted speaker profile $\bar{d}_n$ (Eq. \eqref{eq:asrout}).
Note that $\bar{d}_n$ is computed 
from the speaker inventory $D$
in the speaker identification block, and will be
explained in the next paragraph.
The posterior probability
of token $i$ (i.e. $i$-th token in the dictionary) at the $n$-th decoder step is represented as
 \begin{align}
P(y_n=i|y_{1:n-1},s_{1:n},X,D) \sim o_{n,i}, \label{eq:tokenprob}
\end{align}
where $o_{n,i}$ represents
the $i$-th element of $o_n$.

On the other hand, the speaker identification block is represented as follows.
\begin{align}
 H^{spk} &= \{h^{spk}_1,...,h^{spk}_T\} = {\rm SpeakerEncoder}(X).  \label{eq:spkenc} \\
p_n&=\sum_{t=1}^T \alpha_{n,t}h^{spk}_t, \label{eq:spkvec}\\
q_n &= {\rm SpeakerQueryRNN}(p_n,y_{n-1},q_{n-1}), \label{eq:spkquery}\\
\beta_{n}&= {\rm InventoryAttention}(q_n, D), \label{eq:invatt}\\
\bar{d}_n&=\sum_{k=1}^{K}\beta_{n,k}d_k. \label{eq:weighted_prof}
\end{align}
Firstly, the SpeakerEncoder module converts $X$ into 
a sequence, $H^{spk}$, of 
embeddings representing the speaker features of
the input $X$ (Eq. \eqref{eq:spkenc}).
At every decoder step $n$,
we reuse the attention weight $\alpha_n$ from the ASR block and apply them over
$H^{spk}$
to extract an attention-weighted average, $p_n$, of the speaker embeddings (Eq. \eqref{eq:spkvec}).
The SpeakerQueryRNN module then generates a
speaker query $q_n$ given $p_n$,
the previous output $y_{n-1}$, and the previous 
speaker query $q_{n-1}$ (Eq. \eqref{eq:spkquery}).
Next,
InventoryAttention module 
estimates
attention weight $\beta_n=\{\beta_{n,1},...,\beta_{n,K}\}$ over profiles in $D$
given the speaker query $q_n$ (Eq. \eqref{eq:invatt}).
The attention weight $\beta_{n,k}$ can be seen as 
a posterior probability of person $k$ speaking the $n$-th token
given all the previous tokens and speakers as well as $X$ and $D$, i.e., 
\begin{align}
P(s_n=k|y_{1:n-1},s_{1:n-1},X,D) \sim\beta_{n,k}. \label{eq:spk-prob}
\end{align}
Finally,
 the weighted speaker profile $\bar{d}_n$
 is calculated as the weighted average of the profiles in $D$
 (Eq. \eqref{eq:weighted_prof}).
 As explained earlier, $\bar{d}_n$ is inputted to the ASR block to achieve speaker-biased 
 token estimation (Eq. \eqref{eq:asrout}).

By using Eqs. \eqref{eq:tokenprob} and \eqref{eq:spk-prob}, we can represent 
the joint posterior probability of token $Y$ and speaker $S$ given input $X$ and $D$
as follows.
\begin{align}
P(Y,S|X,D) =&\prod_{n=1}^{N}\{P(y_{n}|y_{1:n-1}, s_{1:n}, X, D) \nonumber \\ 
&\;\;\;\times P(s_{n}|y_{1:n-1}, s_{1:n-1}, X, D)^\gamma \}. \label{eq:samll-2}
\end{align}
Here,
$\gamma$ is 
a scaling parameter 
for the speaker estimation probability 
introduced in \cite{kanda2020joint}.

\subsection{SA-MMI Training}

In \cite{kanda2020joint},
all  network parameters are
optimized with SA-MMI training,
where
the joint posterior probability 
$P(Y,S|X,D)$
is maximized
over training data.
In the form of 
the loss function to be minimized,
SA-MMI training is
represented as follows.
\begin{align}
\mathcal{L}^{\scriptscriptstyle \mathrm{SA-MMI}}&=\sum_r -\log P(Y_r,S_r|X_r,D_r). 
\end{align}
Here, $r$ is a training sample index. Terms $X_r$, $D_r$, $Y_r$, and $S_r$
represent the input speech, speaker inventory, reference token sequence and
reference speaker identity sequence of the $r$-th training sample, respectively.
In the SA-MMI training, we set a scaling parameter $\gamma$ to 0.1 per \cite{kanda2020joint}.

\subsection{Decoding}
\label{sec:decoding}
An extended beam search algorithm is used for decoding with 
the E2E SA-ASR.
In the conventional beam search,
each hypothesis 
contains estimated tokens accompanied by the posterior probability of 
the hypothesis.
In addition to these,
a hypothesis for the proposed method contains
speaker estimation $\beta_{n,k}$.
 Each hypothesis expands until $\langle eos\rangle$ is detected,
 and the estimated tokens in each hypothesis are segmented by  
 $\langle sc\rangle$ 
 to form multiple utterances.
 For each utterance,
  the average of  $\beta_{n,k}$ values, including the last token corresponding to 
   $\langle sc\rangle$
   or $\langle eos\rangle$, 
is calculated for each speaker. 
The speaker with the highest average $\beta_{n,k}$ score is 
 selected as the predicted speaker of that utterance.
 Finally, when the same speaker is predicted for multiple
 utterances, those utterances are concatenated to form a single utterance.
 Note that, in our experiment, we applied length normalization \cite{graves2012sequence} when
 comparing the posterior probability of the hypotheses in the beam.
 Namely, we used the normalized score $P(Y,S|X,D)^{1/|Y|}$ 
 for beam search, where $|Y|$ is the length of sequence $Y$.

\section{SA-MBR Training}
\label{sec:sambr-training}

In this paper, we propose to train the E2E SA-ASR model parameters 
by minimizing the expected SA-WER over training data.
The proposed loss function to be minimized is represented as follows.
\begin{align}
&\mathcal{L}^{\scriptscriptstyle \mathrm{SA-MBR}}=\sum_r \bar{\mathcal{E}_r}, 
\end{align}
where $\bar{\mathcal{E}_r}$ is the expected number of errors based on SA-WER calculation for $r$-th training sample:
\begin{align}
&\bar{\mathcal{E}_r}=\sum_{\scriptscriptstyle Y,S\in \atop\mathcal{B}(X_r,D_r)} \hat{P}(Y,S|X_r,D_r) \mathcal{E}(Y,S; Y_r,S_r).  
\end{align}
Here, $\mathcal{B(X,D)}$ represents 
the $N$-best hypotheses obtained by the extended beam search 
(described in Section \ref{sec:decoding})
given input audio $X$ and speaker inventory $D$.
    The function $\mathcal{E}$ computes the 
number of errors in hypotheses $\{Y,S\}$ 
given reference $\{Y_r,S_r\}$
according to the error counting of SA-WER calculation.
Specifically, we calculate the edit distance between the hypothesis and the reference of
each speaker and sum them up over
all speakers appearing in $S$ and $S_r$.
%Namely, we 
$\hat{P}(Y,S|X,D)$ is a normalized posterior over the $N$-best hypotheses as,
\begin{align}
\hat{P}(Y,S|X_r,D_r)=
\frac{P(Y,S|X_r,D_r)^{1/|Y|}}{\sum\limits_{\scriptscriptstyle Y',S' \in \atop\mathcal{B}(X_r,D_r)} P(Y',S'|X_r,D_r)^{1/|Y'|}},
\end{align}
where $P(Y,S|X_r,D_r)$ is computed by Eq. \eqref{eq:samll-2} with $\gamma=1.0$.
Note that, as we do in the beam search, we apply
the length normalization for each raw posterior 
to compute the normalized posterior.

For each $N$-best hypotheses $\{\tilde{Y},\tilde{S}\}\in\mathcal{B}(X_r,D_r)$,
the error w.r.t. $o_{n,i}$ is calculated as follows
if and only if the estimated token at the $n$-th position of $\tilde{Y}$ is token $i$.
\begin{align}
\frac{\delta \mathcal{L}^{\scriptscriptstyle \mathrm{SA-MBR}}}{\delta \log(o_{n,i})}=
\frac{1}{|\tilde{Y}|}\hat{P}(\tilde{Y},\tilde{S}|X_r,D_r)\{\mathcal{E}(\tilde{Y},\tilde{S};Y_r,S_r)-\bar{\mathcal{E}_r}\}. \nonumber
\end{align}
Otherwise, the error w.r.t. $o_{n,i}$ is zero.
The error w.r.t. $\beta_{n,k}$ is calculated by using exactly the same expression. Namely,
if and only if the estimated speaker at the $n$-th position of $\tilde{S}$ is speaker $k$,
\begin{align}
        \frac{\delta \mathcal{L}^{\scriptscriptstyle \mathrm{SA-MBR}}}{\delta \log(\beta_{n,k})}=
\frac{1}{|\tilde{Y}|}\hat{P}(\tilde{Y},\tilde{S}|X_r,D_r)\{\mathcal{E}(\tilde{Y},\tilde{S};Y_r,S_r)-\bar{\mathcal{E}_r}\}. \nonumber
\end{align}
Otherwise, the error w.r.t. $\beta_{n,k}$ is zero.

In the training, the model parameters are firstly optimized
by the SA-MMI training until they are fully converged.
Then, the well-trained model parameters are further updated by SA-MBR training.
Note that, in the past literature of MBR training,
it was often reported that combining
other training criterion (such as cross entropy criterion)
during MBR-training
improved the accuracy \cite{su2013error,prabhavalkar2018minimum}.
However, we didn't observe any improvement by combining SA-MMI training
with SA-MBR training in our preliminary experiments.

\begin{table*}[t]
  \caption{SER (\%), WER (\%), and SA-WER (\%) for  
E2E SA-ASR trained with SA-MMI and SA-MBR.
The number of profiles per test audio was 8. 
Each profile was extracted by using  2 utterances (15 sec on average). 
No LM was used in the evaluation.}
  \label{tab:overview}
  \centering
{\footnotesize
  \begin{tabular}{c||ccc|ccc|ccc||ccc}
\hline
\multirow{2}{*}{\backslashbox{Training}{Data type}} &  \multicolumn{3}{c|}{1-speaker} & \multicolumn{3}{c|}{2-speaker-mixed } & \multicolumn{3}{c||}{3-speaker-mixed} & \multicolumn{3}{c}{\bf Total}\\
 &  SER & WER & {\bf SA-WER} &  SER & WER & {\bf SA-WER} &  SER & WER & {\bf SA-WER} &   SER &  WER & {\bf SA-WER}  \\
    \hline
\multicolumn{1}{l||}{SA-MMI} & 0.2 & 4.2 & {\bf 4.5}  & 2.5 & 8.6 & {\bf 9.9} & 10.2 & 20.1 & {\bf 23.1} &  6.0 &  13.6 & {\bf 15.6} \\ 
\multicolumn{1}{l||}{SA-MMI $\rightarrow$ SA-MBR} & 0.3 & 4.1  & {\bf 4.5}  & 2.4 & 8.3 & {\bf 9.5} & 9.0  & 18.5 & {\bf 20.7} & 5.3  & 12.7  & {\bf 14.2} \\ 
     \hline
  \end{tabular}
}
  \vspace{-8mm}
\end{table*}

\begin{table}[t]
  \caption{Speaker counting accuracy (\%) 
  before and after SA-MBR training.}
  \label{tab:spk-count}
%  \vspace{-3mm}
  \centering
  {\footnotesize
  \begin{tabular}{l|c|cccc}
    \hline
& Actual \# of Speakers& \multicolumn{4}{c}{Estimated \# of Speakers (\%)} \\
& in Test Data & 1 & 2 & 3 & $>$4 \\
    \hline
& 1 & {\bf 99.96} & 0.04 & 0.00 & 0.00\\ 
SA-MMI & 2 & 2.56 & {\bf 97.44} & 0.00 & 0.00 \\
& 3 & 0.34 & 24.92 & {\bf 74.73} & 0.00\\ 
    \hline
& 1 & {\bf 99.96} & 0.04 & 0.00 & 0.00\\ 
$\rightarrow$ SA-MBR & 2 & 2.29 & {\bf 97.71} & 0.00 & 0.00 \\
& 3 & 0.42 & 21.22 & {\bf 78.32} & 0.04\\ 
    \hline
  \end{tabular}
  }
  \vspace{-6mm}
\end{table}

\begin{table}[t]
  \caption{SA-WER (\%) without or with the length normalization.}
  \label{tab:lennorm}
%  \vspace{-3mm}
  \centering
  {\footnotesize
  \begin{tabular}{c|cc}
    \hline
 Length normalization &  SA-MMI & $\rightarrow$ SA-MBR \\
    \hline
       & 16.1  & 15.7   \\ 
  $\surd$& 15.6  & \bf{14.2}   \\ 
    \hline
  \end{tabular}
  }
  \vspace{-5mm}
\end{table}

\begin{table}[t]
  \caption{SA-WER (\%) 
  with various $N$-best size in SA-MBR training. The beam size in decoding was fixed to 16.}
  \label{tab:training-beam}
%  \vspace{-3mm}
  \centering
  {\footnotesize
  \begin{tabular}{c|cc|c}
    \hline
$N$-best size  &  \multicolumn{2}{c|}{SA-WER (\%)} & Relative \\
 in MBR training &  SA-MMI & $\rightarrow$ SA-MBR & improvement (\%)\\
    \hline
  2  & 15.6  & 14.8  & 5.1 \\ 
  4  & 15.6  & \bf{14.2}  & \bf{9.0} \\ 
  8  & 15.6 & 14.5 & 7.1 \\ 
    \hline
  \end{tabular}
  }
  \vspace{-6mm}
\end{table}

\begin{table}[t]
  \caption{SA-WER (\%) 
  with different beam size in decoding. The $N$-best size in the SA-MBR training was 4.}
  \label{tab:decoding-beam}
%  \vspace{-3mm}
  \centering
  {\footnotesize
  \begin{tabular}{c|cc|c}
    \hline
Beam size  &  \multicolumn{2}{c|}{SA-WER (\%)} & Relative \\
 in decoding &  SA-MMI & $\rightarrow$ SA-MBR & improvement (\%)\\
    \hline
  1  &  16.7 & 15.7  & 6.0 \\ 
  2  &  15.9 & 15.1 & 5.0 \\ 
  4  &  15.6 & 14.5 & 7.1 \\ 
  8  &  15.6 & 14.4  & 7.7  \\ 
  16  &  15.6 & \bf{14.2} & \bf{9.0} \\ 
    \hline
  \end{tabular}
  }
  \vspace{-5mm}
\end{table}

\section{Experiments}
\subsection{Evaluation settings}
\subsubsection{Evaluation data}
We evaluated the effectiveness of the proposed method by using
simulated multi-speaker signals originally from the LibriSpeech corpus~\cite{panayotov2015librispeech}.
Following the Kaldi \cite{povey2011kaldi} recipe, 
we used
the 960 hours of LibriSpeech training data (``train\_960'') for model learning,
the ``dev\_clean'' set for hyper-parameter tuning,
and the ``test\_clean'' set for testing. 

Our training data were generated as follows. 
For each utterance in train\_960, 
randomly chosen $(S-1)$ train\_960 utterances were added after being shifted by random delays, where $S$ was varied from 1 to 3. 
When mixing the audio signals, the original volume of each utterance was kept unchanged, resulting in an average signal-to-interference ratio of about 0 dB. 
As for the delay applied to each utterance, the values were randomly chosen under the constraints 
that (1) the start times of the individual utterances differed by 0.5 sec or longer
and that (2) every utterance in each mixed audio sample had at least one speaker-overlapped region with other utterances.
For each training sample, speaker profiles were generated as follows. 
First, the number of profiles was randomly selected 
from $S$ to 8. 
Among those profiles, $S$ profiles were for the speakers involved in the overlapped speech. 
The utterances for creating the profiles of these speakers were different from those constituting the input overlapped speech. 
The rest of the profiles were randomly extracted from the other speakers in train\_960. 
Each profile was extracted by using 
10 utterances.

The development and evaluation sets were generated from dev\_clean or test\_clean, respectively, in the 
same way as the training set except that 
constraint (1) was not imposed. 
Therefore, 
multiple utterances were allowed to start at the same time
in our evaluation.
Also, each profile was extracted from 
2 utterances (15 sec on average) instead of 10. 
We tuned hyper parameters by using the development set, and report the result on the evaluation set.

\subsubsection{Evaluation metrics}
We evaluated the model with respect to 
 speaker error rate (SER),
WER, and SA-WER. 
{\bf SER} is defined as the total number of  speaker-misattributed utterances generated by the model
 divided by
the number of reference utterances. 
All possible permutations of the hypothesized utterances were examined
by ignoring the ASR results,
and the one that yielded the smallest number of errors
(including the speaker insertion and deletion errors) was picked for the SER calculation. 
Similarly, {\bf WER} was calculated 
by picking the best permutation in terms of the number of word errors (i.e., speaker labels were ignored). 
Finally, {\bf SA-WER} was calculated 
by comparing the ASR hypothesis and the reference transcription of each speaker.
We used SA-WER as the primary evaluation metric.

\subsubsection{Model settings}
In our experiments,
we used a 80-dim log mel filterbank, extracted every 10 msec, for the input feature.
We stacked 3 frames of features and applied the model
to the stacked features.
For the speaker profile, we used
a 128-dim d-vector \cite{variani2014deep}, 
whose extractor was separately 
trained on VoxCeleb Corpus \cite{nagrani2017voxceleb,chung2018voxceleb2}.
The d-vector extractor 
consisted of 17 convolution layers followed by an average pooling layer, which was a modified version of the one presented in \cite{zhou2019cnn}. 

The AsrEncoder consisted of
 5 layers of 1024-dim 
bidirectional long short-term memory (BLSTM), interleaved with layer normalization~\cite{ba2016layer}.
The DecoderRNN consisted of 
2 layers of 1024-dim unidirectional LSTM,
and the DecoderOut consisted of 1 layer of 1024-dim unidirectional LSTM. 
We used a conventional location-aware content-based attention \cite{chorowski2015attention} with 
a single attention head. 
The SpeakerEncoder had the same architecture as the d-vector extractor except for not having the final average pooling layer.
The SpeakerQueryRNN consisted of 1 layer of 512-dim unidirectional LSTM.
We used 16k subwords based on a unigram language model \cite{kudo2018subword}
as a recognition unit. 
We applied volume perturbation to the mixed audio to
increase the training data variability.
Note that
we applied neither 
an additional language model (LM) nor
any other forms of 
data augmentation 
for simplicity. 

As explained in Section \ref{sec:sambr-training}, we firstly
optimized the model parameters based on the SA-MMI training until the model
was fully converged.
The SA-MMI training was performed by using exactly the same settings reported
in \cite{kanda2020joint}.
All
parameters were updated 
by using
an Adam optimizer
with a learning rate of $2\times10^{-5}$. % 0.00002.
We used 8 GPUs, each of which worked
on 6k frames of minibatch.
We report the results 
of the dev\_clean-based best models found after 160k of training iterations. 
As with \cite{kanda2020investigation},
we initialized the parameters of
 AsrEncoder, Attention, DecoderRNN, and DecoderOut
 by using pre-trained SOT-ASR parameters \cite{kanda2020sot} 
while initializing   
the SpeakerEncoder parameters 
 by using those of the d-vector extractor. 
 
After the SA-MMI training,
we further updated
the model parameters based on the SA-MBR training.
The entire network was updated based on $\mathcal{L}^{\scriptscriptstyle\mathrm{SA-MBR}}$
by using
an Adam optimizer
with a learning rate of $4\times10^{-7}$. 
$N$-best hypotheses were generated on the fly,
and 
we used the $N$-best size of 4 unless otherwise stated.
Each 
 minibatch consisted of 8 samples, and
we report the results 
of the best model for the development set within 20k of training iterations.

\subsection{Evaluation results}
\subsubsection{SA-MMI v.s. SA-MBR}

Table \ref{tab:overview} shows the SER, WER, and SA-WER of the E2E SA-ASR model 
based on the SA-MMI training and SA-MBR training. 
In this experiment, we used the extended beam search with a beam size of 16.
As shown in the table,
we observed 
a 9.0\% relative SA-WER reduction (15.6\% to 14.2\%) in total. 
We observed that the SA-MBR training was
especially effective for the most difficult test case,
i.e. 3-speaker mixed test case.
Because SA-MBR training optimizes the model parameters to reduce {\it total} SA-WER,
it would be reasonable that the accuracy of the most error-prone case was mainly improved.
It is also important that %,
1- and 2-speaker test cases were also improved 
or at least on par with the SA-MMI training.

We also evaluated the impact of SA-MBR training  on 
the speaker counting accuracy. 
The result is shown in Table \ref{tab:spk-count}.
We observed a significant improvement in the speaker counting accuracy for the
3-speaker-mixed case from 74.73\% to 78.32\%.
Note that we didn't apply any heuristics to
improve the speaker counting accuracy of the model.
In SA-MMI training, the mis-recognition of
the speaker change symbol in the transcription is counted as only one error.
However, in SA-MBR training, the mis-recognition of the speaker change symbol
could be more severely penalized since it usually causes a large SA-WER degradation. % by the loss function.
We think this is the reason of
significant improvement of speaker counting by the SA-MBR training.

We also examined the effect of the length normalization, the results of which are shown in Table \ref{tab:lennorm}.
In case of not using the length normalization, we excluded it both in training ($N$-best generation and 
the normalized posterior 
calculation) and decoding.
As shown in the table, 
the length normalization had a critical role to achieve the good improvement by SA-MBR training.
It would be
because our training data had very large variance of sequence length due to the variety of
number of speakers.

\subsubsection{Effect of beam size in training and decoding}

To further analyze the SA-MBR training,
we investigated the effect of the beam size of SA-MBR training and decoding.
Firstly, we evaluated the SA-MBR training with different $N$-best sizes.
For decoding, we fixed the beam size to 16. 
The results are shown in Table \ref{tab:training-beam}.
As can be seen in the table, 
a larger $N$-best size did not necessarily lead to better result,
and the best result was obtained with $N=4$.

We also evaluated the effect of the beam size in decoding.
The result is shown in Table \ref{tab:decoding-beam}.
We used the model trained by SA-MBR with the 4-best hypotheses.
As shown in the table, the improvement of SA-MBR training became
more prominent
when we used a larger beam size for decoding.

\section{Conclusions}

In this paper, we proposed
SA-MBR  training  
where the parameters of the E2E SA-ASR model are
trained to minimize the expected SA-WER over the training data.
The proposed SA-MBR training
achieved 9.0\% of relative SA-WER reduction compared with the SA-MMI model 
in LibriSpeech-based experiments. 

% \newpage
\bibliographystyle{IEEEtran}

\bibliography{mybib}

\end{document}